\documentstyle[prl,aps,multicol,psfig]{revtex}

\begin{document}
\draft

\author{A.\,Yu.~Kuntsevich, D.\,A.~Knyazev, V.\,I.~Kozub$^+$,
V.\,M.~Pudalov, G.~Brunhaler$^*$, G.~Bauer$^*$}
\address{P.\,N.~Lebedev physical institute RAS,
119991 Moscow, Russia\\
$^+$ A.F. Ioffe institute RAS, 194021 St.Petersburg, Russia \\
$^*$Institut f\"{u}r Halbleiterphysik, Johannes Kepler
Universt\"{a}t, Linz, Austria}

\title{Nonmonotonic Temperature Dependence of the
Hall Resistance for 2D Electron System in Si}
\maketitle


\begin{abstract} Weak field Hall resistance  $\rho_{xy}(T)$ of the  2D
electron system in Si was measured over the range of temperatures
 $(1-35)$\,K and densities, where the diagonal resistivity
exhibits a ``metallic'' behavior. The $\rho_{xy}(T)$ dependence
was found to be non-monotonic with a maximum at temperatures
$T_{\rm max} \sim 0.16T_{\rm F}$. The $\rho_{xy}(T)$ variations in
the low-temperature domain ($T<T_{\rm max}$) agree qualitatively
with the semiclassical model, that takes into account a broadening
of the Fermi-distribution solely. The semiclassical result
considerably exceeds an interaction-induced quantum correction. In
the ``high-temperature'' domain ($T>T_{\rm max}$), the
$\rho_{xy}(T)$ dependence can be qualitatively explained in terms
of either a semiclassical $T$-dependence of a transport time, or a
thermal activation of carries from a localized band.
\end{abstract}


\begin{multicols}{2}
Origin of the  ``metallic'' type conduction  ($ d\rho/dT>0$) in
two-dimensional (2D) high-mobility electron system in Si remains
in the focus of researcher's interest for over a decade. Several
models have been suggested to explain the metallic behavior:
temperature-dependent screening of impurity potential
\cite{Dolgopolov}, interaction-induced quantum corrections
\cite{Zala}, macroscopic liquid and solid electron phase
separation \cite{Spivak} etc. Though the theories
\cite{Dolgopolov,Zala,Spivak} are substantially different from
each other, all of them explain $\rho_{xx}(T)$ qualitatively or
even quantitatively. Therefore, additional experimental data are
needed for testing these theoretical models. Such data can be
obtained, in particular, from Hall resistance measurements.

Hall resistance of 2D system in weak fields has been theoretically
studied in
Refs.~\cite{Finkelshtein,Zala2,Mirlin,Altshuler,DasSarma}. As
shown in Ref.~\cite{Finkelshtein}, in the frameworks of the
Fermi-liquid theory, Hall resistance is not renormalized and
remains equal to that for the Fermi-gas as $T\rightarrow 0$. Zala,
Narozhny and Aleiner \cite{Zala2} have calculated a quantum
correction  $\delta \rho_{xy}$ to the Hall resistance due to electron-electron
interaction  for arbitrary $T\tau$ (hereafter we
assume $\hbar=1$, $k_B=1$) :
\begin{equation}
\label{Zala} \frac{\delta \rho_{xy}}
{\rho_{xy}}=\frac{e^2(1+n_tC(F^{\sigma}_0))}{\pi^2\sigma_D}
\ln\left(1+\frac{11\pi}{192}\frac{1}{T\tau}\right).
\end{equation}
Here  $\tau$ is the transport time, $n_t$ ($=15$ for two-valley
system) is a number of triplet terms $C(F^{\sigma}_0)$ which
depend on a Fermi-liquid coupling constant $F^{\sigma}_0$. The
correction is small in a ballistic regime ($T\tau\gg 1$); in a
diffusive regime ($T\tau\ll 1$) it is proportional to $\ln(T\tau)$
converging to the Altshuler-Aharonov result \cite{Altshuler}:
\begin{equation}
\label{Altshuler} \frac{\delta \rho_{xy}}{\rho_{xy}}=2\frac{\delta
\rho_{xx}}{\rho_{xx}}
\end{equation}

In the frameworks of a semiclassical screening theory, Das Sarma
and Hwang has suggested another mechanism for temperature
dependence of the Hall coefficient \cite{DasSarma}:
\begin{equation}
\label{DasSarma1} \rho_{xy}=\frac{H}{nec}\frac{<\tau^2>}{<\tau>^2},
\end{equation}
where $\tau$ is averaged mainly over an interval $T$ around the
Fermi energy. To the leading order in temperature the above
relation depends only on smearing of the Fermi distribution and
equals:
\begin{equation}
\label{DasSarma2}
\frac{<\tau^2>}{<\tau>^2}=1+\frac{\pi^2}{3}(T/T_{\rm F})^2.
\end{equation}
As temperature increases, the $\tau(T)$ dependence becomes of
major importance and causes a non-monotonic behavior of
$\rho_{xy}$ as a function of temperature. It is easy to see, that
the $\rho_{xy}(T)$ dependences, predicted by the theories
\cite{Zala2,DasSarma} differ qualitatively from each other. At the
same time, both theories yield $\rho_{xx}(T)$ in a good agreement
with each other as well as with the experimental data
\cite{DasSarmaQuantCorr,PudalovQuantCorr}. It is noteworthy, the
quantum and semiclassical corrections to the Hall resistance do
not exclude each other and seem to be taken into account
simultaneously.

In the present paper, we report on measurements of the temperature
dependence of the weak field Hall resistance and compare the data
to the theories \cite{Zala2,DasSarma}. The measurements were
carried out in the temperature range 1 to 35\,K. The studied
Si-MOS structure of a Hall bar geometry, 5mm$\times$0.8mm, had a
maximum mobility $\approx 25000$\,cm$^2$/Vs at $T=0.3$\,K.
Magnetic field $B=0.1$\,T applied perpendicular to the 2D-plane
was sufficiently high for measuring the Hall voltage. On the other
hand, this field was small enough to ensure the inequality
$\omega_c$$\tau$$\ll 1$ (in the present work $\omega_c\tau\leq
0.05$ for all measurements ), as needed for the applicability of
the theories \cite{Zala2,DasSarma} and for suppressing
Shubnikov-de Haas (SdH) oscillations.

AC measurements were performed at current $I_x=20$\,nA and
frequency 7.6\,Hz with a lock-in amplifier, that detected both
real and imaginary components of the signal. Hall voltage $V_y$
was measured for two opposite field directions. The results were
averaged to eliminate the admixture of $V_{x}$ to $V_{y}$. At
temperatures $T<1$\,K, the two results, $\rho_{xy}(B)$ and
$|\rho_{xy}(-B)|$,  differed substantially due to a non-zero
imaginary component in $V_y$. Therefore, we analyzed only the data
for temperatures higher than $1$\,K. For each of the field
directions, the sample was slowly heated, while the gate voltage
$V_g$ (and hence the concentration $n$) periodically varied in a
step-like fashion. Two components $\rho_{xx}$ and $\rho_{xy}$ were
measured as functions of temperature for each $V_g$ value. A cycle
of measurements took about 16 hours.

\begin{figure}[ht]
\centerline{\psfig{figure=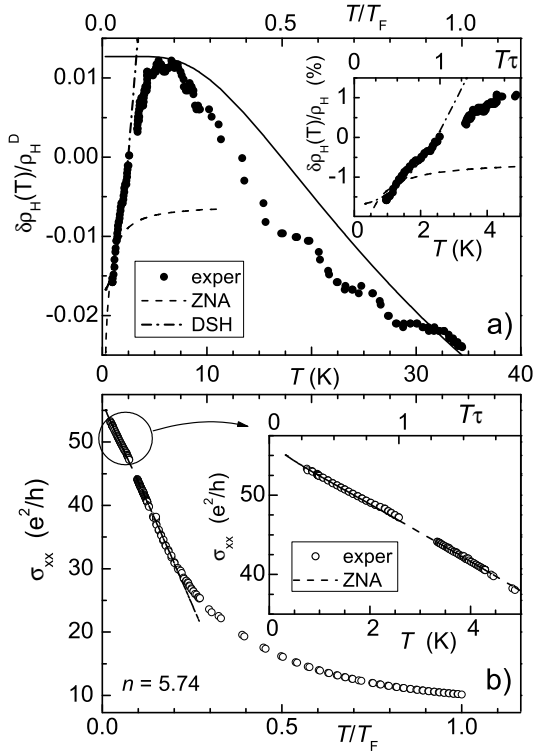,width=230pt}}
\begin{minipage}{3.2in}
\vspace{0.1in}
\caption{Fig.~1. a) Temperature dependence of a deviation of the
Hall resistance from it's classical value (dots). b) Temperature
dependence of the diagonal component of conductivity (empty dots).
Electron density is $n=5.7\times 10^{11} $cm$^{-2}$. The insets
show low-temperature domains of the $\rho_{xy}(T)$ and
$\sigma_{xx}(T)$ dependences. Dashed lines are the calculated
quantum corrections Eq.~(1) \protect\cite{Zala}. Dash-dotted line is
the semiclassical correction Eq.~(\protect\ref{DasSarma2}). Solid
line is the thermal activation calculated according to
Eq.~(\protect\ref{model2}).}
\end{minipage}
\label{fig:F1}
\end{figure}

Figures 1a and 2a show variations of the Hall resistance
$\delta\rho_{xy}$ and the diagonal component of the conductivity
as functions of temperature for electron density $5.7\times
10^{11}$\,cm$^{-2}$ and $11.7\times 10^{11}$\,cm$^{-2}$,
respectively. The variation of the Hall resistance was calculated
with respect to its classical value $H/(n_{\rm SdH}ec)$, where
$n_{\rm SdH}$ was determined from the frequency of SdH
oscillations in a temperature range $T=0.5-2$\,K. There is a well
defined maximum on the temperature dependences of
$\delta\rho_{xy}$; this maximum moves to higher temperatures as
the electron density increases. The maximum variation in $\rho_{xy}$
is about $3\%$. In contrast, $\sigma_{xx}$ decreases monotonically
by a factor of five in the same temperature range (see Figs.~1\,b
and 2\,b).

\begin{figure}[ht]
\centerline{\psfig{figure=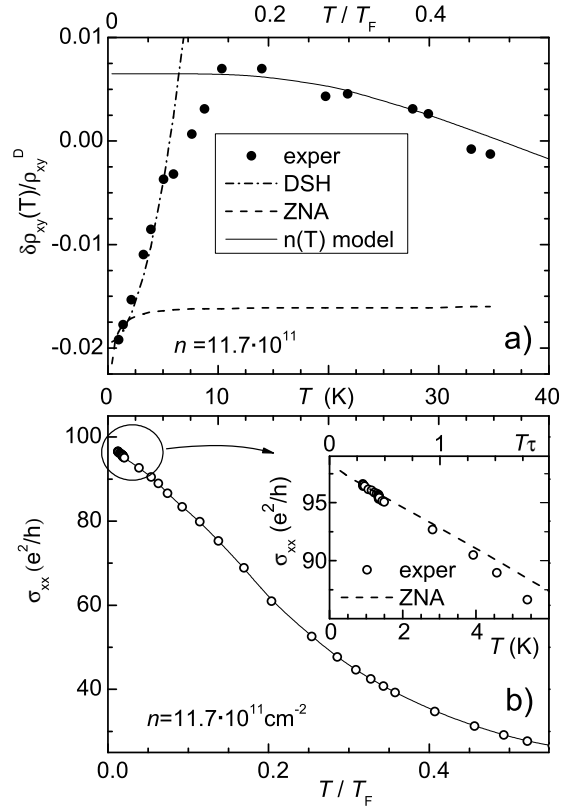,width=220pt}}
\begin{minipage}{3.2in}
\vspace{0.1in}
\caption{Fig.~2. a) Temperature dependence of a deviation of the
Hall resistance from it's classical value (dots). b) Temperature
dependence of the diagonal component of conductivity (empty dots).
Electron density is $n=11.7\times 10^{11} $cm$^{-2}$. The inset
shows low-temperature domain of $\sigma_{xx}(T)$. Dashed lines are
the calculated quantum corrections Eq.~(1) \protect\cite{Zala}.
Dash-dotted line is the semiclassical correction
Eq.~(\protect\ref{DasSarma2}). Solid line is the thermal
activation calculated according to Eq.~(\protect\ref{model2}).}
\label{fig:F2}
\end{minipage}
\end{figure}

{\bf ``Low-temperature'' domain} $(T \ll T_F$,\ $T\tau \lesssim
1)$. The quantum corrections to the diagonal conductivity
\cite{Zala} quantitatively describe the low-temperature part of
the $\sigma_{xx}(T)$ dependence with no adjusting parameters
(dashed lines in Figs. 1b,\,2b). For calculating this correction,
we used independently determined $m^*$ and $F_0^{\sigma}$ values
\cite{Pudalov}.

The quantum correction to $\rho_{xy}$ was evaluated using
Eq.~(\ref{Zala}) with the same $m^*$ and $F_0^{\sigma}$ (see
dashed lines in Figs. 1a and 2a); it appears to be an order of
magnitude less than the experimentally measured
$\delta\rho_{xy}(T)$. At the same time, the semiclassical
dependence Eq.~(\ref{DasSarma2}) (dash-dotted line in Figs. 1a,\,2a)
does agree with the low-temperature part of the
$\delta\rho_{xy}(T)$ dependence. This consistency points at the
leading role of the semiclassical rather than quantum effects in
$\rho_{xy}(T)$ for temperatures $T>1$\,K. It should be noted that
both theoretical curves were arbitrarily shifted vertically for
the best coincidence with the experimental data. Such a shift is
eligible because the absolute value of concentration has been
determined from SdH oscillations with an accuracy of $\sim 1-2$\%.

{\bf ``High-temperature'' domain} ($T\gtrsim 0.3T_{F}$, $T\tau \gg
1$). The measured $\rho_{xy}(T)$ decreases as temperature grows.
The quantum corrections \cite{Zala} in this domain are negligible,
hence the observed effect is presumably of a semiclassical origin.
An analogous decaying $\rho_{xy}(T)$ dependence has been observed
earlier for 2D electron system in Si \cite{pudalov99} and for 2D
hole system in GaAs \cite{Gao,Yasin}. This effect has been
qualitatively explained in Ref.~\cite{DasSarma} on the basis of
Eq.~(\ref{DasSarma1}) by numerical calculation of the
temperature-dependent $\tau$. This mechanism of the
$\rho_{xy}(T)$-dependence might be of significance for our system
as well. However, such a calculation would require a number of
microscopic parameters of the disorder potential which are
poorly-known and therefore should be used as adjustable ones.

\begin{figure}[ht]
\vskip.05in
\centerline{\psfig{figure=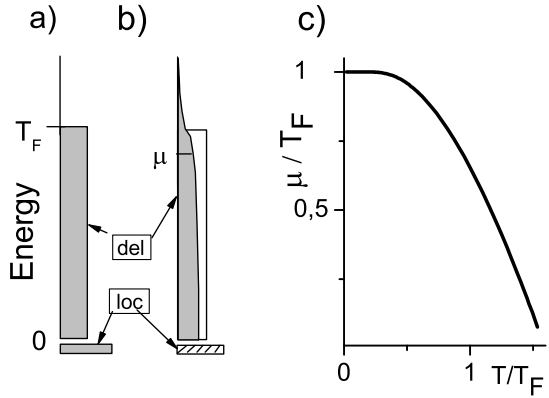,width=230pt}}
\begin{minipage}{3.2in}
\caption{Fig.~3. Schematics of the themoactivation
process for the model with delocalized (del) and  localized (loc)
bands. a) Energy distribution of the electrons at $T=0$. b) The
same for finite $T\sim T_{\rm F}$. Partially filled band of
localized states is hatched; filled areas are shadowed.
c)Temperature dependence of the chemical potential.}
\label{fig:F3}
\end{minipage}
\end{figure}
At higher temperatures $T\sim T_{\rm F}$ another mechanism of the
$\rho_{xy}(T)$ dependence may come into play if the system
contains localized electrons besides the delocalized ones. In this
``two-band'' model, the total electron concentration $n_{\rm tot}$
is independent of temperature and given by the gate voltage. Due
to thermal activation from the localized band, the density of the
delocalized carriers becomes dependent on temperature. As
temperature increases, a number of the delocalized electrons
increases, therefore the Hall resistance decreases.

Let us consider a simple model of this phenomenon, which is
schematically shown in Fig. 3. Let the electron system consists of
the Fermi-gas of the delocalized particles and a localized band
``tail'' \cite{Mott}. The latter has $n_{\rm loc}^0$ states
located in the vicinity of the bottom of the conduction band
\cite{Kozub}, with density of states being $n_{\rm loc}^0\delta
(\varepsilon)$. The total electron concentration is
temperature-independent:
\begin{equation}
\label{model1} n_{\rm tot}=\frac{n_{\rm
loc}^0}{1+e^{-\mu/T}}+\frac{2m}{\pi\hbar^2}T\ln(1+e^{\mu/T}),
\end{equation}
where $\mu$ is the chemical potential. Only the delocalized
carriers contributes to the Hall resistance; their concentration
$n_{\rm del}$  is as follows
\begin{equation}
\label{model2} n_{\rm del}=n_{\rm tot}-\frac{n_{\rm
loc}^0}{1+e^{-\mu/T}}
\end{equation}
In the frameworks of this simple model, there are three parameters
$m$, $n_{\rm tot}$ and $n_{\rm loc}^0$ which determine temperature
dependences of $\mu$ and, consequently, of $\rho_{xy}(T)=H/(n_{\rm
del}ec)$. The first two parameters can be determined
independently, whereas the last parameter is an adjustable one. In
the studied range of carrier densities $n>4\times 10^{11}
$cm$^{-2}$ the renormalization of the effective mass is
insignificant \cite{Pudalov}, therefore we use a bare band mass
$m=0.205m_e$. The variations of $\rho_{xy}$ are small (Fig. 1à,
2à), therefore $n_{\rm tot}\approx n_{\rm SdH}$. The remaining
adjusting parameter $n_{\rm loc}$  for simplicity was chosen
independent of $n_{\rm tot}$.

It appears that $n_{\rm loc}^0=0.7\times 10^{11}$cm$^{-2}$, as
obtained from our fitting, provides a good agreement of the model with
the ``high-temperature'' parts of $\rho_{xy}(T)$ data for various
concentrations (see Figs. 1a and 2a). Is is noteworthy that the
results of calculations plotted in Figs. 1a and 2a were
arbitrarily shifted vertically by 1-2\% for the best coincidence
with experimental data, similar to that it was done in the above
comparison with Refs. \cite{Zala,DasSarma}. It should be
emphasized that $n_{\rm loc}^0$ is approximately equal to $0.5n_c$
for our 2D electron system, where $n_c\approx 1\times
10^{11}$cm$^{-2}$ is the critical density of the metal-insulator
transition (MIT) at $B=0$ \cite{Kravchenko}. This $n_{\rm loc}^0$
value qualitatively agrees with the localized band model
\cite{Kozub} and with ``a few electrons per ion scenario'' of the
MIT \cite{Klapwijk}. Indeed, for a 2D system with bare random
potential, the electrons starts filling  the conduction band only
after the localized electrons have (nonlinearly) screened  the
potential fluctuations.

In conclusion, we  observed a weak ($\sim 2\%$) and
non-monotonic temperature dependence of the Hall resistance for
the  2D electron system in
Si inversion layer, with
a maximum at $T_{\rm max}\approx (0.15-0.20)T_{\rm F}$. In the
same range of temperatures and concentrations, the diagonal
resistivity exhibits strong and monotonic metallic-type behavior.
The low-temperature ($T<T_{\rm max}$) domain of the $\rho_{xy}(T)$
dependence agrees better with the simple semiclassical model
\cite{DasSarma} ($\propto (T/T_{\rm F})^2$), than with the
interaction-induced quantum corrections \cite{Zala2}; this
indicates  that temperature in the studied domain ($T\tau>0.4$) is
too high to observe the quantum corrections. In a
``high-temperature'' domain ($T>T_{\rm max}$), the $\rho_{xy}(T)$
dependence can be explained in terms of either a temperature-dependent
impurity screening \cite{DasSarma}, or a thermal activation of the
localized electrons to the conductivity band. All considered above models do
not contradict to each other and should possibly be taken into
account simultaneously. It is noteworthy that if the concentration
of localized carriers remains about the same down to lower
densities, the localized band can play a significant role in the
metal-insulator transition.

The authors are thankful to G.\,M.~Minkov, A.\,D.~Mirlin and
B.\,N.~Narozhny for discussions. The work was partially supported
by grants from RFBR, the Programs of RAS, the Presidential
support of the leading scientific schools, and the Austrian grant
\#FWF P16160.

\end{multicols}
\end{document}